\documentclass{webofc}
\usepackage[varg]{txfonts}   
\wocname{New Frontiers in Physics 2014}
\woctitle{New Frontiers in Physics 2014}
\begin{document}
\title{STAR Highlights on Heavy Ion Physics}

\author{Shusu Shi\inst{1}\fnsep\thanks{\email{shishusu@gmail.com}}  for the STAR collaboration}

\institute{Nuclear Science Division, Lawrence Berkeley National Laboratory, Berkeley, CA 94720, USA}

\abstract{RHIC-STAR is a mid-rapidity collider experiment for studying high energy nuclear collisions. 
The main physics goals of STAR experiment are 1) studying the properties of the strongly coupled Quark Gluon Plasma 
2) explore the QCD phase diagram structure. In these proceedings, we will review the recent results of heavy ion physics at STAR.
}
\maketitle
\section{Introduction}
\label{intro}
The experimental observations indicate the strongly coupled Quark Gluon Plasma (QGP) has been built up 
in the top energy heavy ion collsions at Relativstic Heavy Ion Collider (RHIC)~\cite{rhicwp1, rhicwp2}. With large acceptance and 
excellent particle identifiaction capbiltiy, the STAR experiment at RHIC is one of best mid-rapidty collider 
experiments to study the properties of new form of matter, QGP, the initial conditions in heavy ion collisions and the 
stucture of the Quantum Chromodynamics (QCD) phase diagram. 
STAR experiment has covered the beam energies of $\sqrt{s_{NN}}$ = 7.7, 11.5, 14.5, 19.6, 27, 39, 62.4 and 200 GeV. 
In the top energy heavy ion collisions (Au + Au at $\sqrt{s_{NN}}$ = 200 GeV), we focus on studying the nature of QGP. 
At the lower collision energy region, the main motivation is to explore the nuclear matter phase structure in the higher 
net-baryon region. 
The extracted baryonic chemical potental ($\mu_{B}$) range based on a statistical model~\cite{statmodel1, statmodel2} from the $0-5\%$ central collisions is 
$20~\leq~\mu_{B}~\leq~420~{\rm MeV}$ which covers a wide region of the QCD phase diagram. In these proceedings, 
we are going to highlight a few results from the recent measurements of the STAR experiment.

\section{Results on azimuthal anisotropy}
\label{sec-1}

The azimuthal anisotropy in the momentum space of final particles relative to the reaction plane is one of the most informative
ways to study the properties of matter created in high energy heavy ion collisions. The coefficients $v_n$ from a Fourier-series 
expansion are used to characterize the event anisotropy quantitatively, which could be represented by the equation 
$v_{n} = \left \langle\cos[n(\phi_{i} - \Psi_{\rm RP})]\right \rangle$, where $\phi_{i}$ means the azimuth of the $i^{\rm th}$ particle in an event, 
$\Psi_{\rm RP}$ means azimuth of the reaction plane and the angle brackets mean an average over all particles in all events. 
$v_1$ is referred to as directed flow, and $v_2$ as elliptic flow.

\subsection{Directed flow}
\label{sec-2}

\begin{figure}
\centering
\sidecaption
\includegraphics[width=8cm,clip]{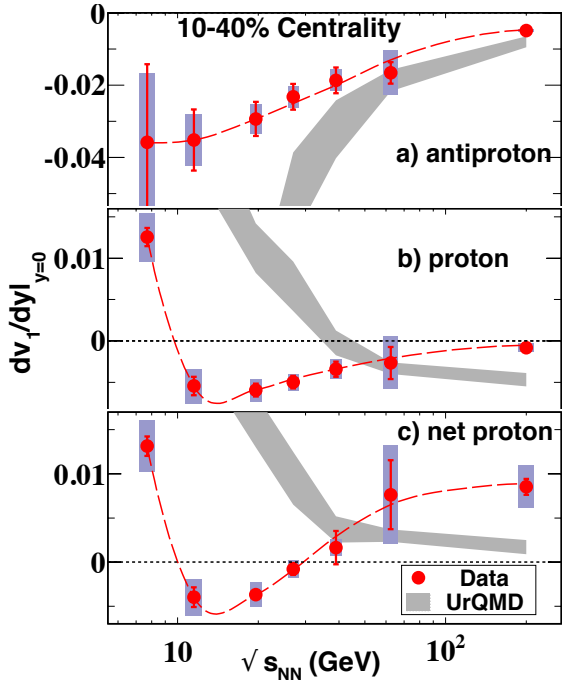}
\caption{The beam energy dependence of directed flow slope ($dv_{1}/dy$) near midrapidity for $10-40\%$ Au + Au
collisions. Panel (a), (b) and (c) report measurement for anti-protons, protons and net-protons respectively, along with UrQMD calculations. 
Shaded bars represent the systematic uncertainties. Dashed curves are a smooth fit to guide the eyes. The figure is taken from~\cite{directedflow}.}
\label{fig-1}      
\end{figure}

The hydrodynamic and nuclear transport models suggest the $v_1$ as a function of rapidity ($y$) in the midrapidity region
offers sensitivity to expansion of the participant matter during the early stage of collisions. Further, a 3-fluid hydrodynamic
calculation with a first-order phase transition between hadronic matter and a QGP predicts a minimum in $dv_{1}/dy$ slope
of net-baryon near midrapidity at $\sqrt{s_{NN}} \sim$  4 GeV, and this minimum has been termed the "softest point collapse"~\cite{v1theory}.
It motivates the energy dependence of directed flow measument. Figure~\ref{fig-1} shows the directed flow slope in
$10-40\%$ Au + Au collisions for protons, anti-protons and net-protons.  The energy dependence of proton $dv_{1}/dy$ involves an
interplay between the directed flow of protons associated with baryon number transported from the initial beam
rapidity to the vicinity of midrapidity, and the directed flow of protons from particle-antiparticle pairs produced near
midrapidity.  Thus, we propose the use of antiproton directed flow as a proxy for the directed flow of produced
protons, and propose that the net-proton slope brings us a step closer to isolating the contributions from transported
initial-state bayonic matter, as well as closer to the net-bayon hydrodynamic calculation. The detailed definition of net-proton slope
could be found in~\cite{directedflow}. In panel (c) of Figure~\ref{fig-1} , it shows the $dv_{1}/dy$ slope of net-proton crosses
zero between 27 and 39 GeV and remains positive up to 200 GeV, where the UrQMD model shows a monotonic trend. The observed beam energy of the minimum in $dv_{1}/dy$ slope
is higher than the energy of the minimum in the hydrodynamic prediction. The observation of net-proton $dv_{1}/dy$ slope is
qualitatively consistent with the predicted signature of a first-order phase transition between hadronic and deconfined matter. 
But more study on other final-state interaction effects, such as annihilation, is necessary to understand the measurement.

\subsection{Elliptic flow}
\label{sec-3}

\begin{figure}
\centering
\sidecaption
\includegraphics[width=9cm,clip]{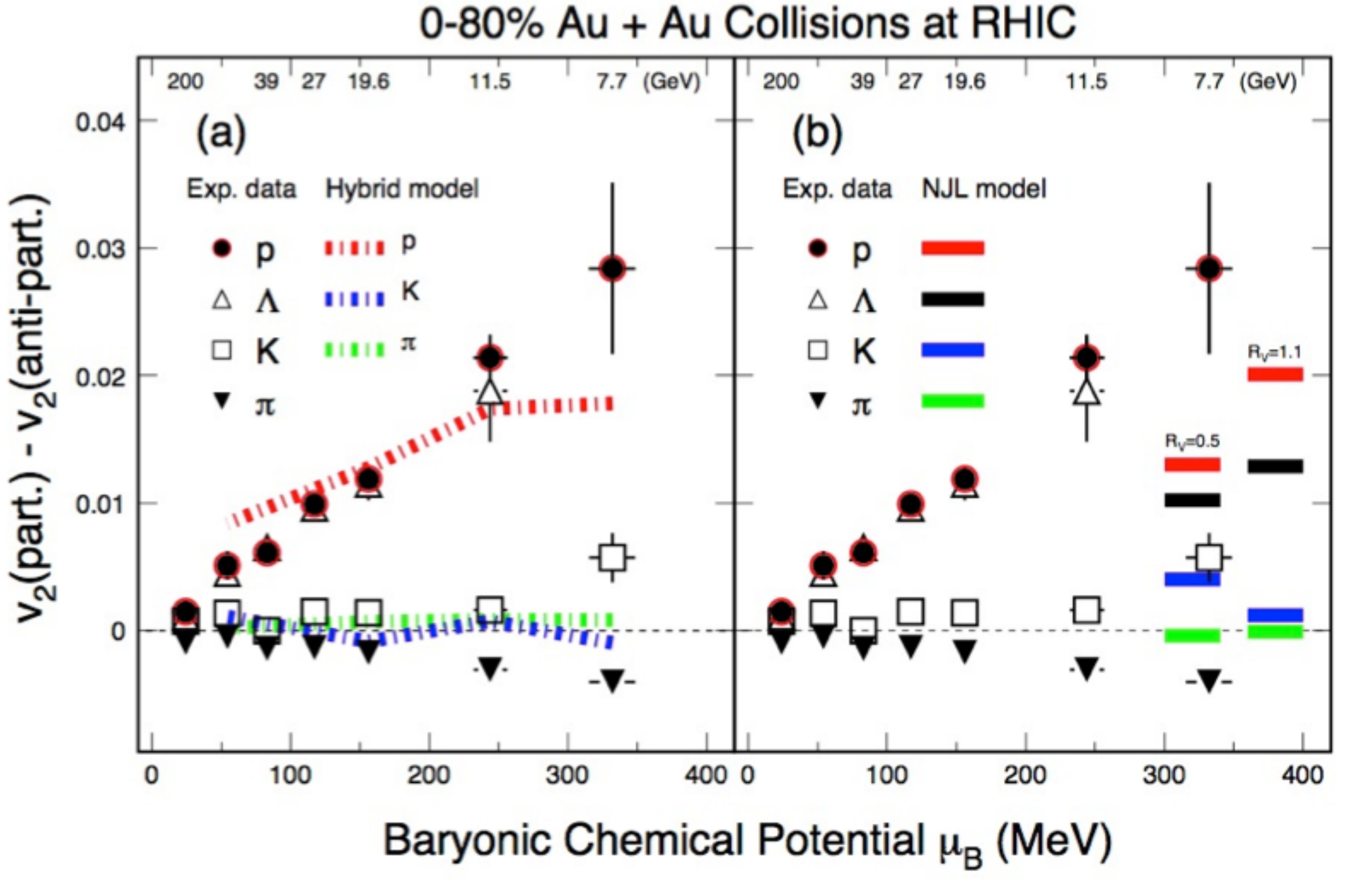}
\caption{The difference in $v_2$ between particles and their corresponding antiparticles as a function of 
baryonic chemical potential for $0-80\%$ central Au + Au collisions from experimental data and theoretical calculations. }
\label{fig-2}       
\end{figure}

The most striking feature on the $v_2$ measurement is the observation of an energy dependent difference in $v_2$
between particles and their corresponding antiparticles~\cite{{ellipticflow}}.
This difference naturally breaks the number of consitutent quark scaling (NCQ) in $v_2$ which is 
regarded as an evidence of partonic collectivity in the top energy heavy ion collisions at RHIC. It indicates the hadronic degrees of
freedom play a more important role at lower collision energies.
Figure~\ref{fig-2} shows the difference in $v_2$ between particles and their corresponding antiparticles as a function of 
baryonic chemical potential. Our data are compared to hydrodynamics + transport (UrQMD) hybrid model~\cite{hybrid} and Nambu-Jona-Lasino (NJL) model~\cite{NJL} which considers both partonic and hadronic potential. The hydrid model could reproduce the baryon (proton) data, but fails to explain the mesons; whereas the NJL model could qualitatively reproduce the hadron splitting. By tuning the $R_v$ parameter which is related to the partonic potential, NJL model fails to reproduce the magnitude for all hadron species simultaneously. 

\begin{figure}
\centering
\sidecaption
\includegraphics[width=9cm,clip]{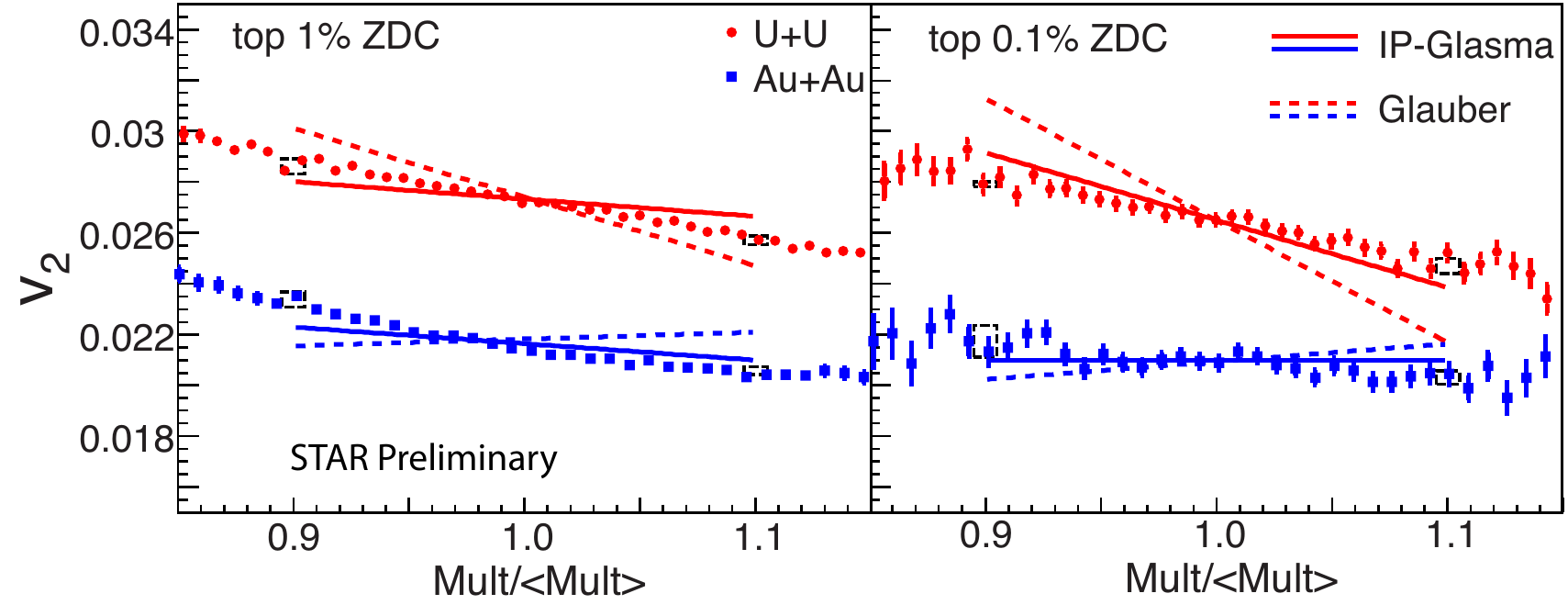}
\caption{The $v_2$ of charged particles as a function of the normalized multiplicity with $\mid\eta \mid < 1$.
The left panel shows the results for the top $1\%$ ZDC central events, while the right panel shows the results for the top
$0.1\%$ ZDC central events. The lines represent Glauber and IP-Glasma simulation slopes calculated from the 
eccentricity. The small boxes indicate systematic uncertainties due to the efficiency corections on the $x$-axis.}
\label{fig-3}      
\end{figure}

Collisions between prolate Uranium nuclei are used to study how particle production and azimuthal anisotropies depend on the
initial geometry in heavy ion collisions. If body-body collisions do produce smaller multiplicities than tip-tip collisions, then we 
expect to see a negative slope in $v_2$ vs multiplicity in fully overlapping U + U collisions. A negative slope, however, can also 
be generated in these collisions by off-axis collisions with larger impact parameters that pass our selection criteria. To assess the 
contribution from this effect, we also study Au + Au collisions as a control sample since Au nuclei are almost spherical. 
Figure~\ref{fig-3} shows the $v_2$ of all charged particles as a function of the normalized multiplicity with $\mid\eta \mid < 1$ 
for two different systems. The left panel shows the results for the 1\% most central events based on the smallest signal seen in the  Zero Degree Calorimeter (ZDCs). 
Both ZDCs are used in the centrality determination. Both Au + Au and U + U show a strong negative slope, which indicates the effect of 
the impact parameter is still prominent. The right panel of Fig. 3 shows the same approach for the $0.1\%$ most central events based on the ZDCs. 
The negative slope for Au + Au collision is smaller in magnitude, indicating the effects from non-central collisions are reduced and 
the variation in multiplicity is mainly driven by fluctuations. For U + U collisions, however, the slope becomes more negative 
as the centrality selection is tightened. This demonstrates that the variation of multiplicity in the $0.1\%$ U + U collisions is dominated 
by the different geometries made possible by the prolate shape of the Uranium nucleus.
We also compare the data to expectation from Glauber model and IP-Glasma calculations. It is clear that the IP-Glasma calculation based on 
gluon saturation describes the data better.

\section{Results on higher moments}
\label{sec-4}

\begin{figure}
\centering
\sidecaption
\includegraphics[width=8cm,clip]{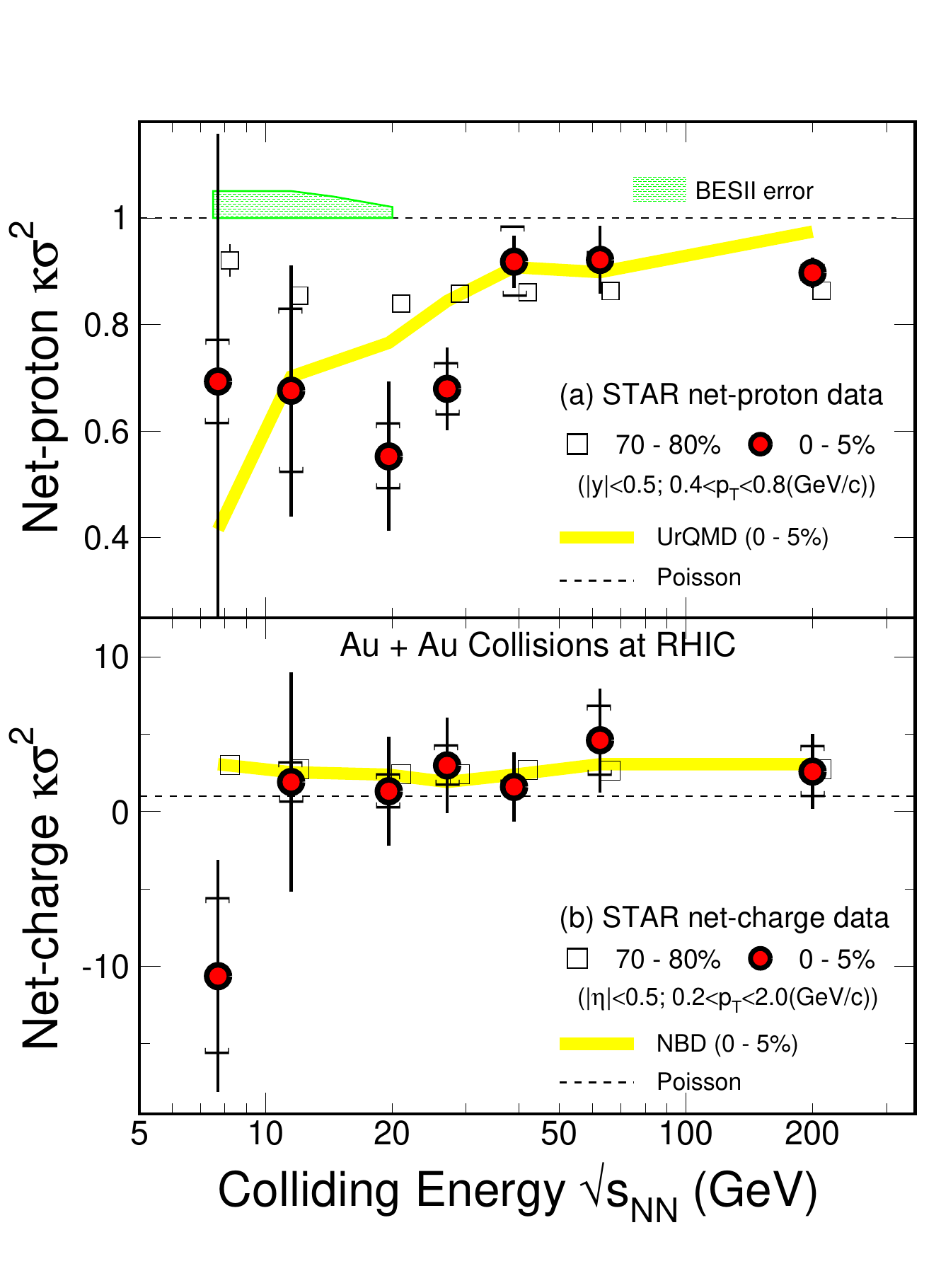}
\caption{Net-proton (top panel)~\cite{netp2014}  and net-charge (bottom
panel)~\cite{netq2014} higher moments $\kappa \sigma^2$ as a function of beam energy from Au + Au collisions at RHIC. The red solid circles 
correspond to $0-5\%$ central collisions and the open squares represent $70-80\%$ peripheral collisions. The vertical 
error bars are statistical and the caps represent systematic uncertainties. The yellow solid line in the top panel
represents $0-5\%$ central Au + Au collision results from UrQMD simulations and the yellow solid line in the bottom 
panel is the result of negative binomial statistics. The dashed lines in both panels represent the Poisson statistics.
The green box in the top panel indicates the estimated statistical errors for net-protons from the RHIC BESII program. }
\label{fig-4}      
\end{figure}

It was argued the large fluctuations of event-by-event multiplicity distributions of conserved quantities such as net-charge, 
net-baryon number, and net-strangeness could be the experimental signature of the QCD critical point. 
Theoretical studies indicate that higher order moments have stronger dependences on correlation length ($\xi$) than 
the variance ($\langle (\delta N)^{3} \rangle$ $\sim$ $\xi^{4.5}$ and $\langle (\delta N)^{4} \rangle$ $\sim$ $\xi^{7}$). 
Thus higher moments have higher sensitivity to the QCD critical point~\cite{mstephanov0911, masakawa09}.
This motivates the study on the kurtosis ($\kappa$) of net-proton (a proxy for net-baryon) and net-charge distributions 
to search for the QCD critical point at STAR. Figure~\ref{fig-4} shows the kurtosis times variance ($\kappa$$\sigma^2$) for 
net-proton (top panel)~\cite{netp2014} and net-charge (bottom panel)~\cite{netq2014} distributions at mid-rapidity 
in Au + Au collisions as a function of colliding energy for $0-5\%$ and $70-80\%$ collision centralities.  
The net-proton $\kappa$$\sigma^2$ values for the 0-5\% centrality selection at $\sqrt{s_{NN}}$ = 19.6 and 27 GeV 
show a relatively larger deviation from Poisson and hadron resonance gas expectation values  (which would correspond to uncorrelated emission and are close to unity) 
and the $\kappa$$\sigma^2$ values from 70-80\% peripheral collisions.
The peripheral collisions are expected to create small systems and do not show significant bulk properties.
The calculations of UrQMD which is a pure hadronic model that does not consider a phase transition show a monotonic
behavior. Our data at lower energies are with large uncertainties, but a possible non-monotonic variation of the $\kappa$$\sigma^2$ of 
the net-proton distribution is not excluded. High statistical dataset for
Au + Au collisions $<$ 20 GeV from the future second phase of the beam energy scan (BESII) will help answer the question. 
In the top panel, the hatched band shows the projection of the statistical errors from BESII.

\section{Results on Chiral Magnetic Effect}
\label{sec-4}

\begin{figure}
\centering
\sidecaption
\includegraphics[width=9cm,clip]{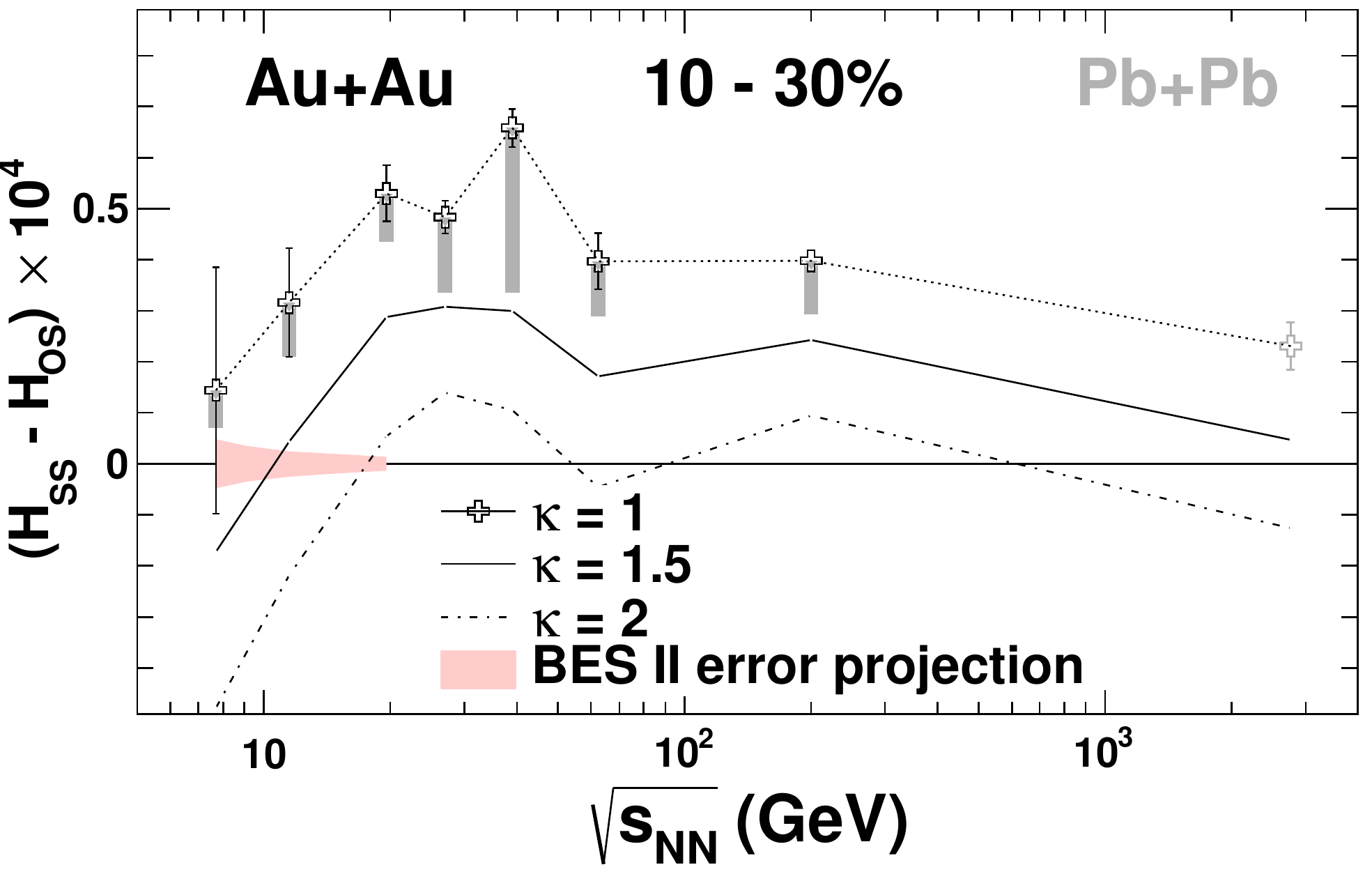}
\caption{Charge separation as a function of collision energy in $10-30\%$ Au + Au collisions. The default values (dotted curve) are 
from $H^{\kappa =1}$, and the solid (dash-dot) curve are obtained with $\kappa$ = 1.5 ($\kappa$ = 2). 
For comparison, the results for Pb + Pb collisions at $\sqrt{s_{NN}}$ = 2.76 TeV are also shown \cite{acme1}.  The vertical asymmetric bands represent 
the systematic errors and the band around zero indicates the statistical errors from the proposed RHIC BESII program. }
\label{fig-5}      
\end{figure}

The RHIC provides a good opportunity to study the parity-odd domains where the local imbalance of chirality results from the
interplay of these topological configurations with the QGP. In heavy ion collisions, energetic spectator protons produce a magnetic field. 
This strong magnetic field coupled with the chirality asymmetry in the parity-odd domains, induces a separation of electric 
charge along the direction of the magnetic field. STAR has proposed a three particle correlator method to study the Chiral Magnetic Effect (CME)~\cite{scme1}. 
Figure~\ref{fig-5} shows the collision energy dependence of the charge separation after subtraction of the background related to  collective flow~\cite{cmebes}. The results approach to zero when the beam energy is below 11 GeV. In order to draw
a clear conclusion, more statistics are needed for the lower beam energies ($<$ 20 GeV). The colored band around zero shows the estimated
statistical errors for the results from RHIC BESII program.

\section{Results on heavy flavor production}
\label{sec-5}

Charm and beauty ($c$ and $b$) quarks are created predominantly via initial hard scatterings in nucleon nucleon colisions and 
the production rate is calculable with perturbative QCD techniques. The large masses are expected to be retained during the 
interactions with the medium~\cite{hf1, hf2}. Therefore the heavy flavor production is a useful tool to study the medium properties
of the early stages of the system when QGP is expected to exist.

\begin{figure}
\centering
\sidecaption
\includegraphics[width=8cm,clip]{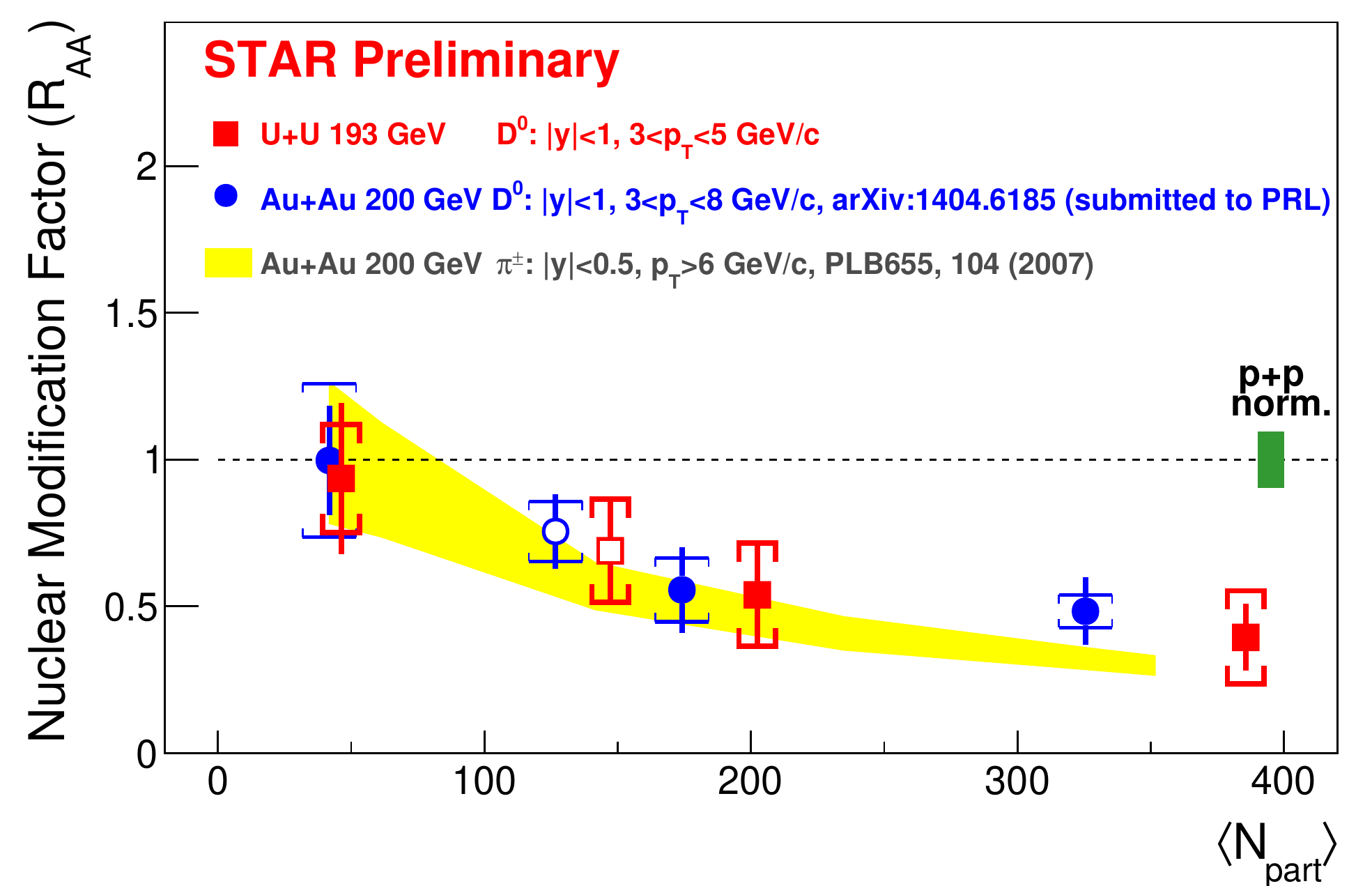}
\caption{Nuclear modification factor ($R_{AA}$) as a function of the number of participant nucleons $\left \langle N_{\rm part} \right \rangle$ 
in Au + Au collisions at $\sqrt{s_{NN}}$ = 200 GeV and U + U collisions at $\sqrt{s_{NN}}$ = 193 GeV. The vertical bars on data
points indicate statistical uncertainties, while the brackets for bin-to-bin systematic uncertainties. 
The size of the box around $R_{AA}$ =1 correspond to global normalization uncertainty. }
\label{fig-6}      
\end{figure}

\begin{figure}
\centering
\sidecaption
\includegraphics[width=8cm,clip]{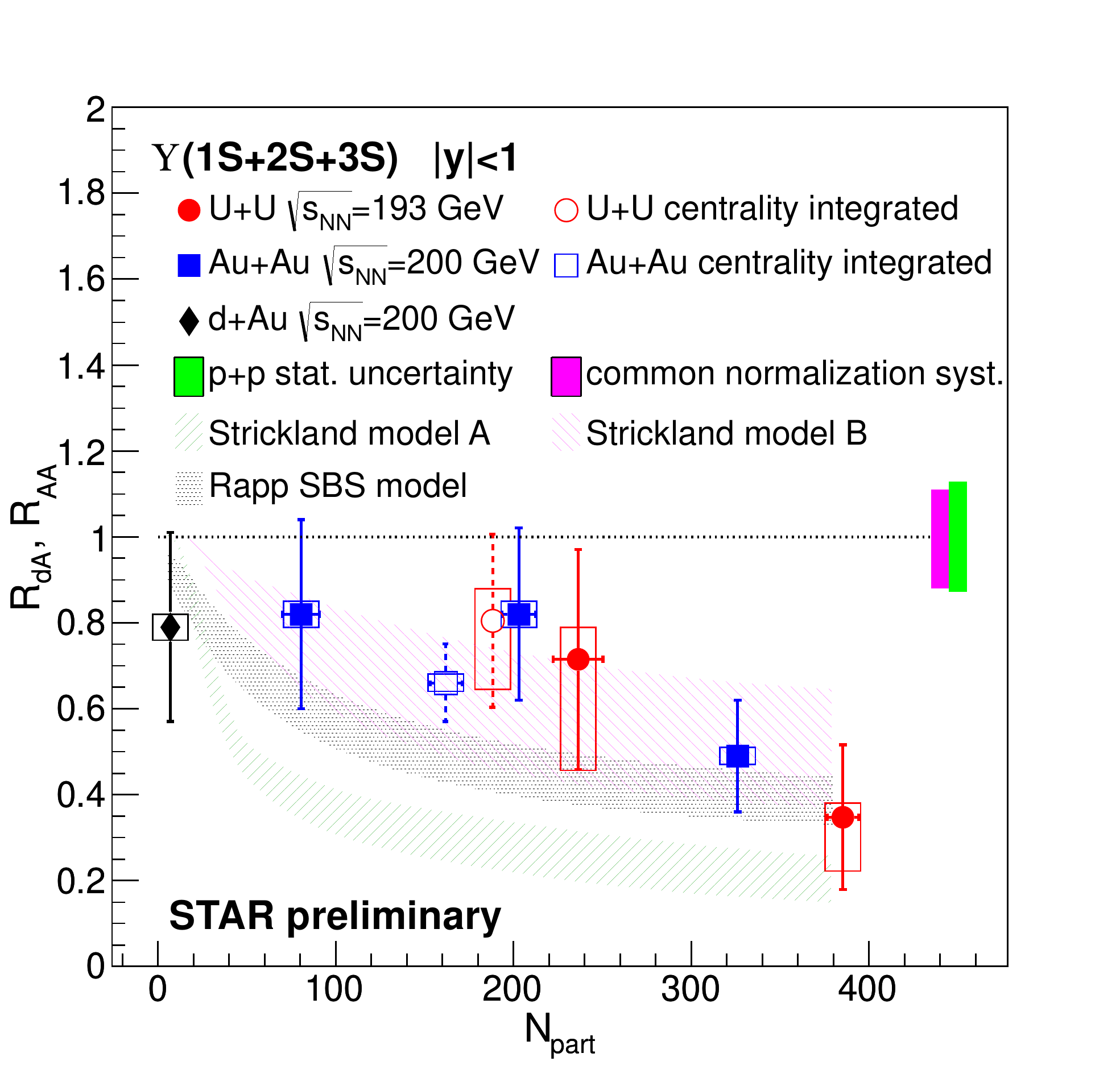}
\caption{The $\Upsilon$(1S+2S+3S) $R_{AA}$ as a function of $\left \langle N_{\rm part} \right \rangle$ in U + U collisions at 
$\sqrt{s_{NN}}$ = 193 GeV, compared to the results in Au + Au, d + Au collisions at $\sqrt{s_{NN}}$ = 200 GeV and model calculations.}
\label{fig-7}      
\end{figure}

Figure~\ref{fig-6} shows the nuclear modifcation factor $R_{AA}$ of $D^{0}$ as a function of the number of participant nucleons $\left \langle N_{\rm part} \right \rangle$.  
We observe the suppression of $D^{0}$ meson production at large $p_T$ in Au + Au and U + U collisions follows a global trend as a function of $\left \langle N_{\rm part} \right \rangle$. Compared to the light flavor hadrons ($\pi^{\pm}$),  $D^{0}$ mesons suffer similar amount of energy loss. It is consistent with physics picture of a substantial amount of charm-medium interactions.
Figure~\ref{fig-7} shows the $R_{AA}$ of $\Upsilon$(1S+2S+3S) as a function of $\left \langle N_{\rm part} \right \rangle$ in
U + U collisions at $\sqrt{s_{NN}}$ = 193 GeV and Au + Au, $d$ + Au collisions at $\sqrt{s_{NN}}$ = 200 GeV.  A significant suppression is observed in central collision events of Au + Au and U + U. The results disfavor Strickland model A, which is based
on the heavy quark free energy scenario. The model of Strickland and Bazow incorporating lattice QCD results on screening and broadening of bottomonium with a potential based on heavy quark internal energy is consistent with the data~\cite{Strickland}. Also, the strong
binding scenario proposed by Emerick, Zhao and Rapp including additionally possible Cold Nuclear Matter (CNM) effects is consistent with our results~\cite{Emerick}.

\section{Results on di-electron production}

\begin{figure}
\centering
\sidecaption
\includegraphics[width=8cm,clip]{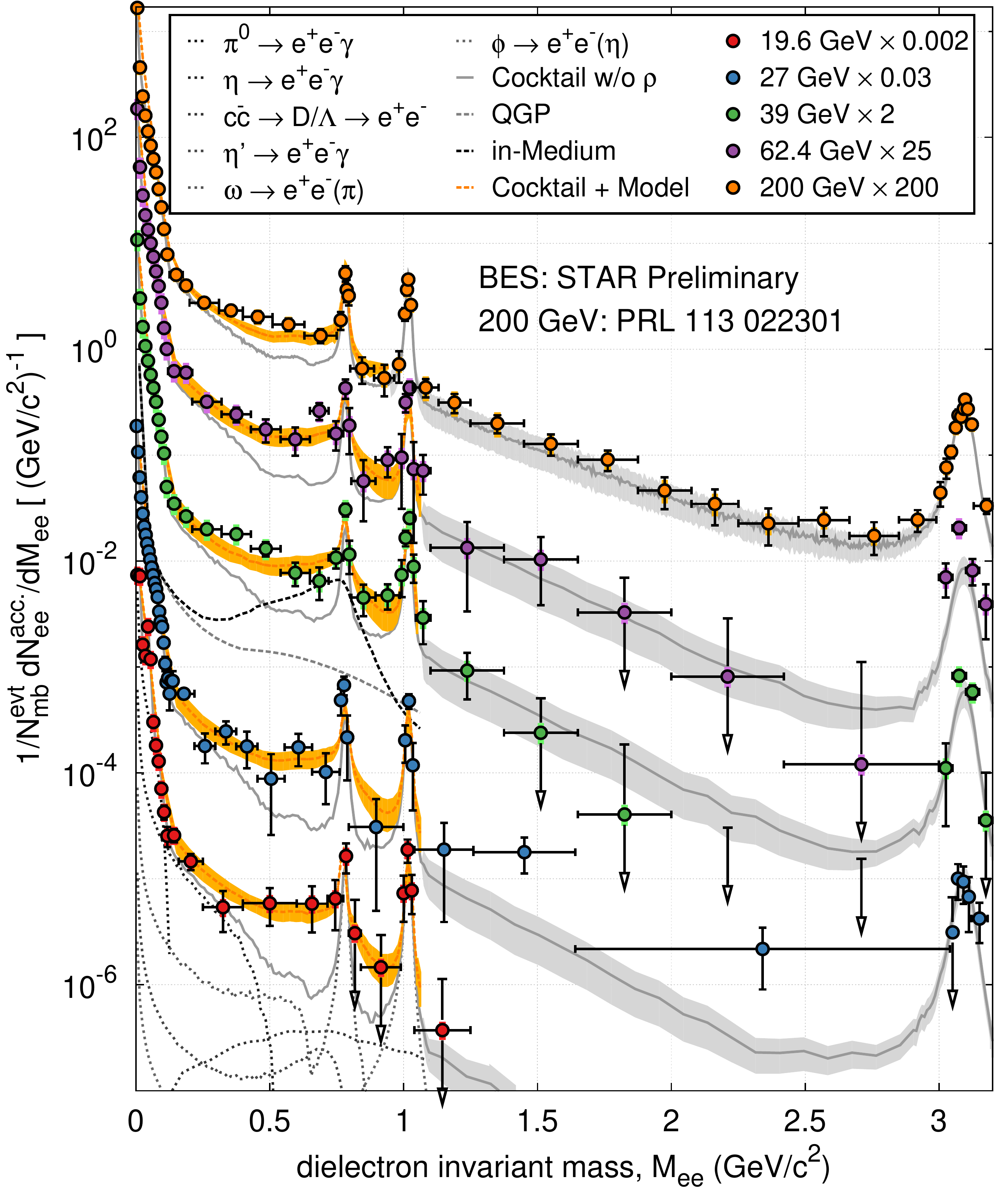}
\caption{The di-electron invariant mass distributions as a function of beam energy in minimun-bias Au + Au collisions at $\sqrt{s_{NN}}$ = 19.6,
27, 39, 62.4 and 200 GeV, from bottom to top, respectively. The yellow bands in the low mass region represent the
results from hadronic cocktails + in-medium model calculations. The gray bands are the systematic uncertainties.}
\label{fig-8}      
\end{figure}

Di-electrons are produced throughout the whole evolution in a heavy ion collision and have negligible final state interactions
with created nuclear matter. Hence, di-electrons can be considered as unique probes to study the hot and dense matter.
Potentially, the di-electron could provide information of chiral dynamics in the low mass region $M_{ll} < 1 $ GeV/$c^{2}$ and
QGP direct radiation in the intermediate mass region $1 < M_{ll} < 3$ GeV/$c^{2}$. Figure~\ref{fig-8} shows the efficiency-corrected di-electron invariant mass distribution from minimum-bias Au + Au collisions at $\sqrt{s_{NN}}$ = 19.6, 17, 39, 62.4 and 200 GeV.
The 200 GeV results are from~\cite{dielectron200}. The results of hadronic cocktails + in-medium model calculations are shown by the yellow bands~\cite{ralf00}. The comparison of model calculations to the invariant mass dependence
of di-electron low mass range yields supports the conclusion that, within experimental uncertainties, in-medium modifications of
the $\rho$ spectral function consistently describe the low mass region enhancement from 19.6 GeV to 200 GeV.

\section{Summary and outlook}

In these proceedings, we highlight the selected STAR heavy ion physics results of azimuthal anisotropy, higher moments, charge separation, heavy flavor production and di-electron production from top energy heavy ion collisions and Beam Energy Scan phase I (BESI). 
The azimuthal anisotropy measurements in the top energy collisions offer us an opportunity to investigate the initial condition
of the collisions.
The heavy flavor production measurements help us to understand the interaction between heavy quark and medium, thus explore the properties of QGP.  The deviation from the QGP signals in the lower energy collisions indicates the hadronic degrees of freedom play a more important role. 

Both Heavy Flavor Tracker (HFT) and Muon Telescope Detector (MTD) were fully installed for
run14 at RHIC in year 2014. The new dataset will allow us to make precise measurement on heavy quark hadron and heavy quarkonium production. With electron cooling plus longer beam bunches for BESII, the luminosity will be improved by 
a factor of $4-15$ compared with BESI. In addition, two related detector upgrades, the Event Plane Detector (EPD) and the inner Time Projection Chamber
(iTPC) will improve the capability of STAR experiment.  BESII program will offer us the unique opportunity to map the predicted
QCD critial point and phase boundary in the phase diagram.

\end{document}